# EmoFit: Affect Monitoring System for Sedentary Jobs

Amol S. Patwardhan, and Gerald M. Knapp

**Abstract**—Emotional and physical well-being at workplace is important for a positive work environment and higher productivity. Jobs such as software programming lead to a sedentary lifestyle and require high interaction with computers. Working at the same job for years can cause a feeling of intellectual stagnation and lack of drive. Many employees experience lack of motivation, mild to extreme depression due to reasons such as aversion towards job responsibilities and incompatibility with coworkers or boss. This research proposed an affect monitoring system EmoFit that would play the role of psychological and physical health trainer. The day to day computer activity and body language was analyzed to detect the physical and emotional well-being of the user. Keystrokes, activity interruptions, eye tracking, facial expressions, body posture and speech were monitored to gauge the user's health. The system also provided activities such as at-desk exercise and stress relief game and motivational quotes in an attempt to promote users well-being. The experimental results and positive feedback from test subjects showed that EmoFit would help improve emotional and physical well-being at jobs that involve significant computer usage.

─────────── ◆ ───────────

## 1 INTRODUCTION

WORKPLACES are second homes to most professionals. A significant part of a person's life is spent at his or her workplace. It is therefore important that the work environment of a person is pleasant. A healthy workplace is important for the emotional and physical well-being of an individual. Conversely, the mental and physical state of individual employees affects the collective team spirit, productivity, client services and customer satisfaction and work culture of a company. This research focused on four key objectives:
1. Tracking the emotional and physical state of technology professionals using data obtained from day to day interaction with computers.
2. Evaluating the tracked data and predicting affective state of the users.
3. Presenting the users with tools to improve mental and physical well-being, in the form of activities that promote positive spirit such as inspirational quotes and games.
4. Implementation of an affect (emotion) detection, evaluation and health improvement tool.

Specifically, these objectives were studied within the context of software programmers, who often spend extended hours working at computers. The majority of programmer's professional time is spent at a desk, leading to a potentially sedentary lifestyle. Various stress factors such as project deadlines, problems with the software in a production environment, pressure to release a fix at short notice, company mergers, economy and company sales and revenue influence the psychological and physical state of individual employees. Repetition of mundane activities for several years can cause an employee to feel uninterested in day to day tasks. These factors can lead to reduced productivity and cause low turnover. By being able to monitor affective state of an employee (happiness, dissatisfaction, anger, frustration, sadness) and physical state (fatigue), management can take preventive or curative measures.

## 2 LITERATURE REVIEW
### 2.1 Organizational Behavior
Research done in [1] discussed the interdependency of affective state of an employee and the overall organization. It showed how emotional state of employees influence the decision making ability, productivity and thought process of a team and individuals working at supervisorial and managerial positions. Study done in [2] examined the relationship between the performance of employees and the feeling of solitude. The study also established that such individuals demonstrated withdrawal from team activities and were uninterested in assigned task which had an effect on efficiency at an organizational level. Research done in [3] provided metrics to selfevaluate emotional state and showed that spreading emotional awareness among individuals helped improve commitment, outcome, unity at the organizational level and eliminated job related insecurities. Researchers [4] showed a relationship between managerial responsibility and tendency to assist others in emotional needs at the workplace. The research also showed that managers with negative affective state did not engage in assisting subordinates. In a study [5] the effects of emotional self-control over performance at workplace and the contribution of managers and senior leadership on employee's mental health was analyzed. The research also concluded that

─────

- *A.S. Patwardhan is with the Department of Mechanical and Industrial Engineering, 2508 Patrick F Taylor Hall, Louisiana State University, Baton Rouge, LA 70803. E-mail: apatwa3@lsu.edu.*
- *G.M. Knapp is with the Department of Mechanical and Industrial Engineering, 2508 Patrick F Taylor Hall, Louisiana State University, Baton Rouge, LA 70803. E-mail: gknapp@lsu.edu.*

interaction between employee and their co-workers had a higher positive dynamics compared to interaction with their supervisor. Work done in [6] provided a psychological insight about the need of giving importance to affect at workplace and the influence of emotion on employee productivity and organization's operation. A research was done [7] on effects of interpreting nonverbal emotional expressions on work environment. It showed that the employees preferred knowledge about positive emotions rather than recognizing negativity through the process of eavesdropping since such information was considered detrimental to workplace. Research done in [8] examined the previous work done on emotion in workplace and provided models and behavioral guide in regulating emotions at workplace. Research work in [9] provided quantitative information about the degree of emotional regulation at workplace and by how much does an employee suppress or express or fake an emotion based on the job role. Research done in [10] analyzed the prediction of emotional state of workers in service industry and use of emotion awareness in improving work environment by reducing stress. Results indicated that the emotional information was essential in coping with work related anguish and recommended use of emotional intelligence in preventive training. The article [11] provided overview on four different papers ranging in various business settings such as internship, electronic job application, selfevaluation of emotion and its application in business and an entrepreneur's emotional response towards a business related development. A study done in [12] evaluated interaction between employee, supervisor and co-workers and the effects of interaction on emotional state. The interaction was analyzed in terms of employee's reaction towards others, reaction by others towards the employee and emotional state of employee while working alone. Results showed that positive emotions were caused by work achievements and rapport with co-worker and supervisor. It also found that past positive emotions neither persisted nor had a direct effect on enhancement in work conditions over long period of time.

## 2.2 Affect in Technology Jobs

This section discusses studies on affective computing at workplace. Researchers conducted surveys [13] among part time young professionals in two separate periods of time and studied the dependency between emotional state and work related activity and decision making across the two time periods. It was found that negative emotions from first reporting duration were directly related to decisions behind leaving the job in the later time period. Study done in [14] evaluated employee behavior in IT industry caused by computer anxiety and showed that managerial support and provision of executing tasks independently helped reduce nervousness. The study also provided models to implement adoption of technology in workplaces without causing emotional unease. Research work done in [15] conducted empirical examination of an individual's attitude towards computers and its effect on job satisfaction, career and workplace environment. Study done in [16] concluded that understanding the individual's behavior was important in increasing the productivity in virtual work environment. An example of a virtual work environment is working remotely from home while logged into office computer over the internet. A study [17] evaluated the attitude and mental focus of computer users and its influence on e-learning. Research was done in [18] to introduce personality and user emotion awareness into a system to improve the end user experience. A model was developed to present the user with a more personalized interaction making the experience of using technology more effective and interesting. The research analyzed video, streaming in the background which was then fed into the affect recognition system to drive a user interface that responded to user's emotional state and behavior. A study [19] analyzed how programmers underwent various emotional phases over several months at the job. The research showed the effects of the programmer's emotional state on their productivity and quality of code.

## 3 OVERVIEW

This research performed a preliminary study on the emotional and physical state of programmers by analyzing day to day activity of software programmers. The research implemented a system that could predict the mental state of its users based on facial expression, keystroke, activity interruption, body posture and speech. The system also incorporated health improvement tools such as interactive game and inspirational quotes. One of the ob-jectives of this research was to implement a health monitoring and improvement system called EmoFit. This system incorporated the three different modules as follows 1) Multimodal Affect Tracking, 2) Affect Predictor, 3) Health Booster Component.

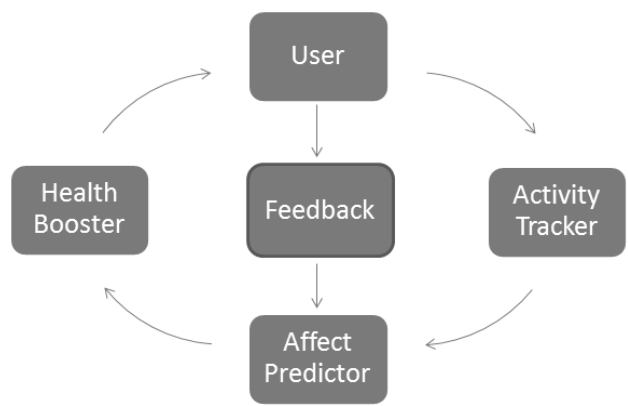

Fig. 1. Conceptual diagram of the EmoFit System and its subcomponents. The arrows indicate the direction of data flow and the interaction between the system and the user.

The conceptual diagram of the EmoFit system shows the flow of data between different components. The user activity is tracked by the Activity Tracker. The tracked data is passed to the Predictor component. The user feedback from questionnaire is also fed into the Predictor component. This data is used to generate reports about emotional and physical state of the programmer. The user can choose to interact with the various health improvement tools from the health booster component such as game, inspirational and funny quotes based on reports available from the Predictor component. The EmoFit system incorporated various functionalities such as keystroke logging, activity interruption, questionnaire to obtain feedback on emotional and physical state, report on mental and physical state, interactive game, quotes and questionnaire on evaluation of the system. Fig. 1 shows the conceptual diagram of the EmoFit system. User provided feedback to the system which was used to validate the predictions from the Affect Predictor. The activity tracker provided the features from various modalities such as keystroke, activity interruption, eye tracking, body posture tracking, facial expression tracking and speech. The prediction results were used to display a report in the form of a bar graph showing the emotions experienced by the user. The health booster component included health improvement tools such as inspirational and funny quotes and game to promote mental and physical wellbeing at the programmer's workplace. Experimental setup consisted of a windows based program called EmoFit installed on the programmer's workstations and a Microsoft Kinect sensor was used for monitoring facial expressions, speech and body posture. The experiments were conducted on 5 individuals working as software programmers in 4 different companies. The EmoFit program received data from the sensor and this data was used for predicting emotional and physical state of the user. Due to the sensitive nature of the data such as the content of the code, emails and messages typed by the programmers during the day and words captured from the speech modality the experiment were conducted under the conditions of anonymity in order to preserve the privacy of the participants. The research not only proposed use of this tool as an emotional and health monitoring and improvement system by individual programmers but also as a performance evaluation system that could be used by the management for identifying degradation in productivity caused by work related emotional and physical health issues. The research identified a possible use of the system by managers attempting to document activity and communication of programmers performing tasks below expectations.

## 4 MULTIMODAL AFFECT TRACKING

The first objective of this research was to track the emotional and physical state of the programmer through day to day computer usage from multiple channels of input. This research used different modalities such as keystroke logging, body posture tracking, eye-ball movement, breaks taken between tasks and speech. Data capturing tools such as Windows based application and Microsoft Kinect were used to record the data while user's performed their day to day tasks. The following sections discuss the details of data captured using various modalities. For this research only workstations having Microsoft Windows operating system were considered.

### 4.1 Keystroke Logging

Software programmers spend considerable amount of time during the office hours typing thousands of lines of code. Keystroke data provides data on the average duration of continuous typing sessions per programmer, the average time spent by programmers between breaks and the frequency of typing sessions and speed of typing. This research examined the relationship between the tracked keystrokes and the user's mental state obtained from the user feedback. The correlation was also useful to estimate the level of interest in the assigned task and enthusiasm of user in performing the task.

The research implemented keystroke logging to capture the frequency of computer usage, duration between successive sessions of typing and words typed per minute during the day. The tracked data was stored in a comma separated text file. The tracking was done in a nonintrusive manner running as a background process with the consent of participating programmers. For the purpose of tracking the words typed by the programmers and the timestamp when each word was input to the system was recorded. Frequency and the keystroke count were used as the metric to establish relationship with the user's emotional state. The keystroke logging took into account the navigational keys such as up, down, left, right, space, tab and return key. The function and control keys such as ctrl, shift, alt and F1 to F12 were also included in the calculation of total number of keystrokes.

### 4.2 Body Posture Tracking

Body language provides important cues about the mental state of a person. For instance, if a person is slouching or walking lethargically then it can be concluded that the person is tired or uninterested in the activity being performed. On the other hand, a programmer sitting upright can be perceived as attentive and content with the task being performed. By tracking the person's body posture valuable data can be obtained to deduce information about the individual's affective state. In this research Microsoft Kinect sensor was used to track the joints of the person. The sensor generates frames containing skeletal data which can be used to record the joints of a human body. The study tracked head, shoulder center, left and right shoulder, center of spine, hip center, left hip, right hip, left elbow and right elbow and both wrists to capture the body posture of a person sitting in front of desk and using a computer. The joint data was available from the Kinect sensor in the form of frames. The x and y coordinates of each of the joints and the movement of each joint for a period of 10 frames was tracked. The x, y coordinates of the joints was used as a feature vector and served as an input to the rule based affect predictor component of the EmoFit system.

### 4.3 Eye Tracking

Important information about the computer user's activity can be obtained by capturing which parts of screen the user tends to look and how often the users look at certain areas of the screen. For instance, frequent staring at the top of the screen can indicate that the user may be using a search engine on internet, or typing a URL in the address bar whereas left to right movement can indicate that the user is reading an article. In this study the eye movement of the computer user was tracked. The location on the screen where the users had a tendency to focus for a period greater than 5 seconds was also tracked. This research also tracked the availability of the eyes and presence of the eyes in front of the computer screen. Tracking data from eyes was unavailable when the user was drowsy or had moved away from the screen or had stepped away from the desk. The duration for which the data was unavailable was also tracked. The research examined the relationship between the eye movement and the user's interest in performing the task and whether prediction about the user's state of mind can be made on the basis of eye movement.

### 4.4 Facial Expression Tracking

Facial expressions are an important modality to infer a person's feelings and mental state. Users can express confusion, satisfaction and happiness or sadness using facial expressions. Software programmer who has spent several hours attempting to debug the code and troubleshooting an issue may exhibit a confused and exhausted facial expression. The programmer may express satisfactory grin or a smile when the code executes flawlessly and the task appears to be on track as per deadline. An uninterested or bored programmer may yawn often or rarely smile. This data from facial expression can provide information about the user's level of interest in performing a task. In this research the facial expression data was tracked. The face detection and tracking application programming interface (API) available in Microsoft Kinect software development kit (SDK) was used to track more than 100 facial feature points. Various locations on the lips, cheeks, eyebrows, eyes, chin and forehead of the user were tracked to detect facial expression. The affect predictor component established a relationship between the detected facial features and the affective state of the user.

### 4.5 Interruption in Work Activity

Employees face interruptions in their task because of a scheduled team meeting, town hall meeting, training, chatting with co-workers, internet surfing, coffee, restroom and smoke breaks. This research measured the duration and frequency of such interruptions and examined the influence of these breaks on the user's subsequent activity and affective state. An employee may come back from a meeting and display signs of stress or may show signs of joy after hearing about a promotion or raise in salary after a meeting with supervisor. Similarly, frequent coffee, restroom or smoke breaks may indicate lack of dedication in the employee's efforts. The research evaluated these scenarios and studied the relationship between these behavioral patterns with the emotional and physical state of the computer users. The work interruption tracking was achieved by logging the computer session locked and unlocked states. This was achieved using the Microsoft Win32 session switch event. A function monitoring the event checked a change in the session state from locked to unlock and vice versa to keep track of user activity interruption.

### 4.6 Speech Tracking

Conversation with co-workers or utterance of words while working on a computer can provide clues about the mental state of an employee. A programmer may be frustrated because the code does not work as expected or he may express satisfaction and achievement by uttering words such as "yes", "awesome" and "great". This study recorded utterance of these words by the subjects under observation and the frequency of these words during an 8-hour work day. The research evaluated the hypothesis that a friendly work environment meant more interactions and positive state of mind whereas negativity among co-workers meant longer periods of silence and shorter conversations. The EmoFit system captured the speech using the speech recognition capabilities made available in Microsoft Speech Recognition API. A dictionary of words was created and each emotion was associated with a set of words related to emotional utterance. This dictionary was saved in an xml format. A sample set for the happy emotion class is listed as follows: great, awesome, superb, wonderful, nice, yes, yeah, perfect, amazing, wow, good, delighted, ebullient, ecstatic, elated, energetic, enthusiastic, euphoric, excited, exhilarated, overjoyed, thrilled, tickled pink, turned on, vibrant, zippy, aglow, buoyant, cheerful, elevated, gleeful, happy, in high spirits, jovial, light-hearted, lively, merry, riding high, sparkling, up, contented, cool, fine, genial, glad, gratified, keen, pleasant, pleased, satisfied, serene, sunny, pleasure, jubilant, exultant, joyous, laugh, smile, cheer, cheese. The EmoFit system matched the detected word and predicted the emotion of the speaker by performing a lookup in the dictionary. The Emofit system used a similar set of words for each of the basic emotions and used the dictionary to predict other basic emotions such as Surprise, Anger, Fear, Disgust and Sadness.

## 5 AFFECT PREDICTORS

The second objective of this study was to evaluate the data obtained from the tracking modules and use it to predict the programmer's affective state. This research implemented tracking day to day interaction of users with the computers in terms of key stroke logging, eye movement, facial expression, body posture, intermediate breaks taken by the user and speech. The study examined if there was a relation between the observed data and the emotion and physical state of the users. The evaluation engine consisted of two components. The first component was called rule based affect predictor (RAP). The RAP evaluated the tracked data and attempted to predict user's emotional state and physical state using emotion templates defined by set of rules. The second component was called feedback based affect predictor (FAP). The FAP was used by the programmers to provide their emotional and physical condition twice during an 8-hour work day, in the form of multiple choice questions. Based on the available feedback from the user the FAP generat-ed a daily emotional and physical status report. The following section will discuss each of the two components in depth.

## 5.1 RAP

The RAP module was based on affect recognition rules that were applied on the tracked data from various modalities. This paper provides a sample list of rules based on tracked data from body and facial input modality. For facial expression tracking the Face API available with the Kinect Software Development Kit (SDK) was used. The Face API provides ability to track more than 100 feature points on the face. In this study the left and right eyebrows, left, top and right points on upper and lower lips, left, top and right points on both the eyelids, left and right cheeks, bottom of chin, left, top and right side of forehead were tracked. The tracked facial feature points were used detect emotions such as Happy, Sad, Surprise, Anger, Disgust and Fear using rules that evaluated distance between the tracked points and the relative location of key points with respective to each other. The tracked points were available from the Kinect sensor. For the body posture tracking the left, center and right shoulder, left and right elbow, left and right wrist, center of spine, left, center and right side of hip and the top of head were tracked. This skeleton data was available from the Kinect sensor. For the eye tracking the center of both eyes, left, top and right of both eye lids were tracked. For the speech based tracking, the Kinect provides speech recognition API. A dictionary file containing a list of words representing emotions was created in xml format. The speech recognition component compared the recognized words against the dictionary file and associated the recognized words with the emotion specified in the dictionary file for those words. Some of the rules used in the RAP module for recognizing emotions from body posture and facial feature data are as follows:

1. If both wrists were held behind the head and the elbows were located above the shoulder, then it implied that the user was contemplating and may be sad.
2. If both wrists were above the elbow and below the shoulder in front of the spine, then the user was actively working on the assigned task and may be content and happy.
3. If both the wrist were below the elbows and the head was tilted either to the left or right, then the user may be sad.
4. If the distance between upper eye lid and lower eye lid was less than a threshold t and the distance between left and right tracked points on the cheek and the lower eyelid was less than a threshold d, then the user was annoyed.
5. If the distance between left and right most tracked point on the lips was more than a threshold t, then the user was smiling and was happy.

## 5.2 FAP

The FAP module presented a questionnaire to the end user to log their emotional state and physical state. Each user provided feedback to the system by answering the questions twice a day. This information was useful on assessing the affective state of the workers on a daily basis. This level of feedback could potentially be used for report generation purpose so that supervisors and man-agers could intervene when an employee experienced work related stress. The data was reliable because it came directly from the person undergoing emotional changes as opposed to data obtained digitally from the various modalities. This data was also useful in validating the prediction results of the RAP module.

Fig. 2. The questionnaire used for obtaining the emotional and physical state of computer user.

Fig. 2 shows the system component that displayed the questions to the programmer. The responses were saved as a comma delimited text file. The report generation component parsed this file to display the emotional and physical state of the programmer based on the feedback in the form of a bar graph. The multiple choice questions used for obtaining the mental and physical state of the computer users were as follows:

1. What is your age group?
2. How many years have you worked at this job?
3. How would you rate your overall mental or emotional health at work?
4. What are the reasons for unhappiness at work?
5. What are the reasons for job satisfaction at work?
6. Which emotions do you experience at work?
7. How do you feel physically at work?

The answers to the above questions provided first-hand information about the computer user's emotional and physical state.

## 6 HEALTH BOOSTER COMPONENT

The third objective of this research was to promote healthy lifestyle at workplace using a tool driven by the outcomes of the affect predictor component. The potential health boosting tools are games, emotional diary, mental exercise, motivational quotes, quizzes, inspirational blogs and RSS Feeds. These tools can be used to inculcate positive feelings in the user's mind and simple physical activities to relieve stress and weariness. In order to evaluate the effectiveness of some of these tools a Kinect based dodge ball game, a widget to display inspirational quotes related to workplace and a widget to display funny programmer quotes were implemented. A widget is a very small utility program that can be placed on the computer desktop. Some examples of commonly used widgets are clock, weather and stock market quotes.

## 6.1 Dodge ball game using Kinect

The first tool used in inducing positive emotions was a dodge ball game using Kinect. The purpose of the game was to encourage the user from getting up from the seat and engage in physical activity. The intention was to involve the user in non-work related activity for a small amount of time so that the user felt refreshed mentally and physically.

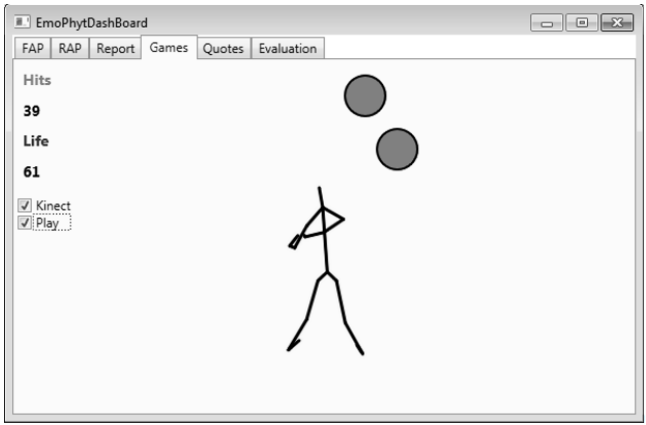

Fig. 3. The purpose of the game was to engage the user in light physical activity by dodging the falling balls.

Many programmers use computer and video games at home and can easily identify with this aspect of technology. As a result, this research implemented a simple game using Kinect with the expectation that it will be readily accepted by the programmers as a tool to promote positive feelings mentally and physically in the workplace. The user had to stand in front of the Kinect sensor and dodge the balls dropping from the top of the screen. This caused them to move their body and served as a physical exercise. Every hit reduced the life by 1 point. If the user managed to not get hit by the ball for 15 consecutive seconds, the life recharged by 1 point. The game was over when the user ran out of lives. The objective of dodging the ball introduced the need for light physical activity.

## 4.2 Quotes Widget

The second tool used in the health booster component was a software widget that displayed inspirational quotes from famous personalities, scientists and writers. A total of 22 quotes were displayed randomly on the screen at an interval of 30 seconds. The third tool used to promote positive mental and physical state was a software widget that displayed funny quotes that programmers could identify. The quotes included references to various programming languages and witty remarks on day to day activities of a typical programmer.

The research implemented this widget with an intention to encourage positive emotions such as laughter, smile and happiness. The widgets also served as a motivational tool. A total of 12 funny quotes were used by the widget for the purpose of the evaluation and displayed every 30 seconds. The funny quotes were intentionally chosen from the field of software programming so that the participants could identify with the funny quotes based on their experience with various programming languages and software development tasks. The quotes for both the widgets were stored in xml format. This provided easy maintainability and additions of more quotes in future.

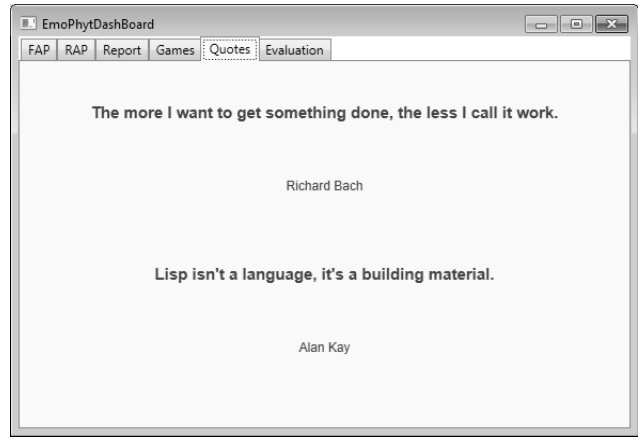

Fig. 4. The inspirational quotes widget and the funny programmer quote widget.

## 7 RESULTS

### 7.1 Keystroke Log

The keystroke log data was analyzed using the affect predictor and provided valuable information about the affective state of the computer users. Among the five programmers who used the EmoFit system, two programmers showed low average typing speed in terms of words per minute. As a result, the total keystrokes logged for those two users during an 8-hour work day were low.

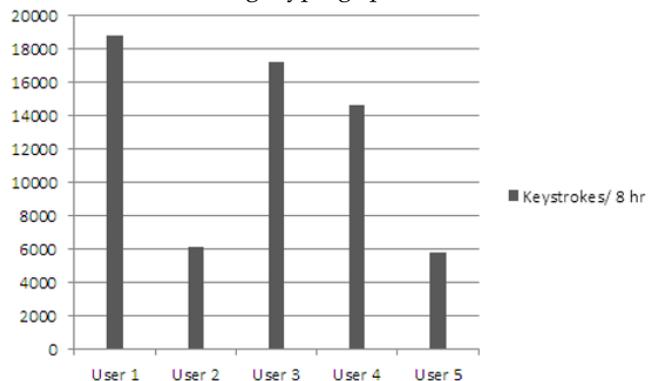

Fig. 5. The keystrokes per 8-hour work day for the five programmers showed low keystroke frequency for the two programmers who reported dissatisfaction from their job.

This was consistent with their feedback about lack of interest in their job and overall low morale. The results showed a clear threshold value of less than 6000 keystrokes per 8 hours for the 2 users with low productivity. On the other hand, the higher keystrokes per 8 hours from the other 3 programmers indicated a positive emotional and physical state and higher productivity in terms of lines of codes typed during the day.

### 7.2 Activity Interruption

The activity interruption was recorded on the basis of how often the user took breaks from the programming activity. Since there was no way to determine the reason for the break, the results were based only on the data about when the user locked the workstation and walked away from the desk. Two programmers took higher number of breaks from the computer usage compared to the other three programmers who displayed overall posi-tive emotional and physical state. The results show a threshold of 10 or more breaks for users who reported lack of interest in the job and poor emotional and physical well-being. Results indicated that the three programmers who reported job satisfaction and happy state of mind showed less than 10 breaks.

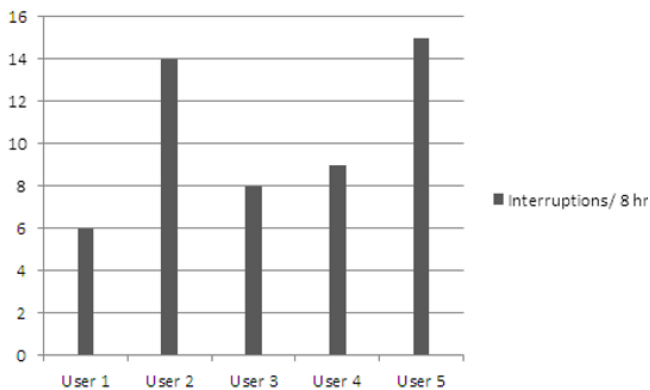

Fig. 6. The activity interruption results indicated higher breaks for 2 out of 5 programmers.

### 7.3 Affect Predictors

The body posture, facial expressions and speech modalities were used to detect the participating programmer's emotions. Each session consisted of 45 minutes of programmer activity in front of the Kinect sensor located at a distance of 1.5 meters from where the user sat. The sessions were repeated twice during the 8-hour session. The results showed high detection rate for happy and surprise emotions for the three users who reported positive emotional and physical health at workplace. This was because the users exhibited positive facial expressions while working and demonstrated more movement of hands and body while performing the programming tasks.

On the other hand, the results showed high detection rate for neutral and sad emotion in case of the two programmers who had reported lack of interest and low morale in the feedback questionnaire. This was because the two programmers showed less activity and neutral facial expressions as compared to enthusiasm and positive body language demonstrated by the other three programmers who were happy with the tasks assigned to them. The results based on the rule based predictor component were in agreement with the feedback provided by the participating programmers. The results were based on majority voting from the results of the individual modalities. For instance, if the facial expression and speech modality detected happy as the emotion and the body modality detected neutral emotion then the overall emotion was chosen to be happy based on two out of three votes from the multiple modalities. Additionally, the emotions were also ranked by the empirical probability. In the above case the empirical probability of happy and neutral emotion was 2/3 and 1/3 respectively. This enabled the RAP to predict multiple emotions for the same user. The predictions were performed using rules on data from 10 consecutive frames at a time.

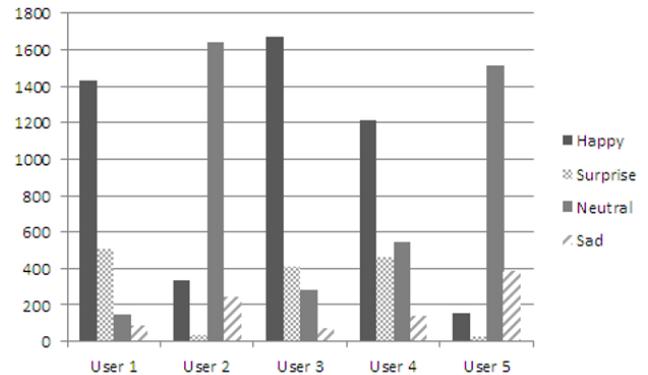

Fig. 7. Affect prediction results from the rule based affect predictor component showing high detection for happiness and surprise among programmers who reported job satisfaction.

### 7.4 Eye Tracking

The results of eye tracking were not conclusive. For all the five users the x and y co-ordinates of the eye gaze did not provide sufficient information to accurately predict the user's affective state. The normalized x-coordinate values of each user's eye tracking data for a period of 40 seconds has been shown in Fig. 8. There wasn't a specific area of screen that showed more attention from any particular user. The x-co-ordinate plot showed that the users viewed various parts of the screen for executing their tasks and no conclusive evidence could be found about whether the user was looking at a web site URL, menu portion of the interactive development environment used for coding or at any internet messaging program. The plot of the xco-ordinate did not show any specific area of the screen to which the users focused their attention. The movements of the eye gaze in the x or y direction did not provide any specific pattern that could be associated with the user's affective state. As a result, any rules to detect affect based on the location and movement of the tracked eye data could not be formulated.

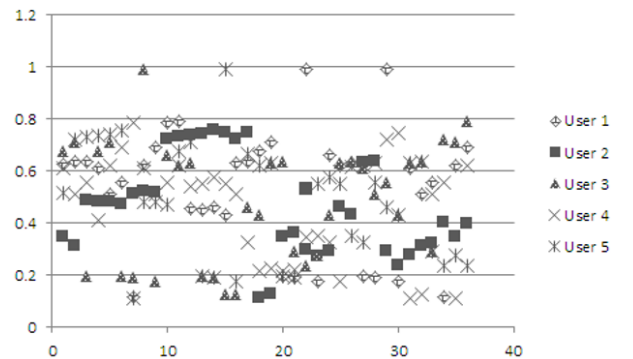

Fig. 8. The scatter plot of x-coordinate of eye tracking data for a period of 40 seconds did not show location or movement specific pattern.

## 7.5 Evaluation

The EmoFit system received a positive feedback from the participating software programmers. All the participants experienced an improvement in their overall emotional and physical state while working during the day and using the health monitoring system as a tool for promoting mental and physical well-being. The EmoFit system in-corporated a questionnaire as an evaluation of the systems effectiveness in detecting emotional state. The questionnaire also evaluated the effectiveness of the system as a tool for stimulating positive feelings in the mind and body of the computer user. The questions used for the evaluation were as follows:
1. What is your age group?
2. How many years have you worked at this job?
3. Did the tracking and survey help assess your emotional and physical state?
4. Did the health booster tools and game make you feel positive?
5. Was the EmoFit System effective in improving emotional and physical positivity at work?

Fig. 9. Evaluation Screen for user feedback of EmoFit system.

The questionnaire for evaluation was integrated into the EmoFit system. Each question was provided with three choices ranging from not effective, somewhat effective and very effective. After the 8-hour monitoring session, the users were asked to submit their evaluations on the effectiveness of the Emofit system. The results from the evaluation on the system by the five programmers are shown in Fig. 10. The participants gave positive feedback about the EmoFit system. The participants also agreed that the system made them feel more positive even if they were already content with their job. None of the participants responded with not effective in their responses.

Majority of the participants found the health improvement tools very effective and found the physical activity involved in the game and the motivational quotes very useful in promoting positive attitude towards their work. Since programmers are interested in video games, the participants liked the idea of using the game as a health enhancing tool and as a medium to refresh themselves mentally after lengthy sessions working on complicated software code.

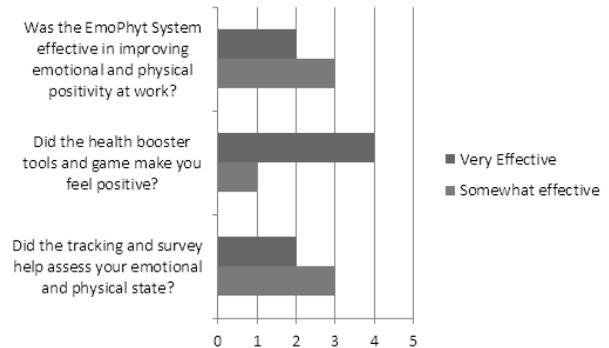

Fig. 10. The results based on user evaluation about the effectiveness of EmoFit system.

The programmers also provided positive feedback about the use of funny programmer quotes because they could easily identify with the humor. The programmers also found the inspirational quotes from well-known personalities as an effective tool to stay motivated.

## 8 CONCLUSION

This research demonstrated how an affect detection, evaluation and health improvement software tool can be effectively used to keep track of individual's emotional and physical state. The tools can also be used to promote positive spirit and well-being in user's mind, especially for IT workers who lead a sedentary lifestyle at work. Due to the technology savvy nature of the job such productivity tools will be readily accepted by the employees without any privacy or intrusion issues because these employees have better understand of the techniques used in capturing data and the type of data obtained.

This study showed that users could be encouraged to engage in physical and stress relief activity, by means of a health assessment tool for workplace such as the EmoFit system. After using the tool, the end users demonstrated more enthusiasm in subsequent activities and a better understanding of their physical and emotional state. EmoFit included intellectual and physical workout options such as dodge ball game, motivational and funny quotes to instill optimistic atti-tude. In addition to the above options the research proposes use of mood based music, display of family photos and motivational emails to be incorporated into the system, as part of future improvements and analysis. These options would assist in inducing constructive emotions among users and help improve the team spirit at workplace.

**Amol S. Patwardhan** received the MS in systems science degree from Louisiana State University in 2006. Currently he is a graduate student in the Department of Mechanical and Industrial Engineering pursuing a PhD in Engineering Science with Information Technology Concentration.

**Gerald M. Knapp** received the MS degree in industrial engineering from the State University of New York at Buffalo and the PhD degree in industrial engineering from University of Iowa. He is currently Fred B. and Ruth B. Zigler Associate Professor in the Department of Mechanical and Industrial Engineering at Louisiana State University.